\documentclass[review,number,sort&compress, 5p]{elsarticle}

\usepackage{natbib}
\usepackage{amsmath}
\usepackage{amssymb}
\usepackage{epsfig}

\journal{Nuclear Instruments and Methods in Physics Research Section A}

\begin{document}

\begin{frontmatter}

\title{Test of candidate light distributors for the muon $(g$\,$-$\,$2)$ laser calibration system}

\author[1,3]{A.~Anastasi}
\author[1]{D.~Babusci}
\author[2]{ F.~Baffigi}
\author[4,10]{G.~Cantatore}
\author[4,12]{D.~Cauz}
\author[1]{G.~Corradi}
\author[1]{S.~Dabagov}
\author[6]{G.~Di Sciascio}
\author[5,13]{R.~Di Stefano}
\author[1,2]{ C.~Ferrari}
\author[15]{A.T.~Fienberg}
\author[1,2]{A.~Fioretti}
\author[2]{L.~Fulgentini}
\author[1,2]{ C.~Gabbanini\corref{cor2}}
\author[2]{ L.A.~Gizzi}
\author[1]{D.~Hampai}
\author[15]{D.W.~Hertzog}
\author[5,11]{M.~Iacovacci}
\author[4,14]{M.~Karuza}
\author[15]{J.~ Kaspar}
\author[2]{P.~Koester}
\author[2]{L.~Labate}
\author[9]{S.~Mastroianni}
\author[6]{D.~Moricciani}
\author[4,12]{G.~Pauletta}
\author[4,12]{L.~Santi}
\author[1]{G.~Venanzoni}
 \address[1]{Laboratori Nazionali Frascati dell' INFN, Via E. Fermi 40, 00044 Frascati, Italy }
\address[2]{Istituto Nazionale di Ottica del C.N.R., UOS Pisa, via Moruzzi 1, 56124, Pisa, Italy }
\address[3]{Dipartimento di Fisica e di Scienze della Terra dell’Universit\`a di Messina, Messina, Italy}
\address[4]{INFN, Sezione di Trieste e G.C. di Udine, Italy}
\address[5]{INFN, Sezione di Napoli, Italy}
\address[6]{INFN, Sezione di Roma Tor Vergata, Roma, Italy}
\address[10]{Universit\`a di Trieste, Trieste, Italy}
\address[11]{Universit\`a di Napoli, Napoli, Italy}
\address[12]{Universit\`a di Udine, Udine, Italy}
\address[13]{Universit\`a di Cassino, Cassino, Italy}
\address[14]{University of Rijeka, Rijeka, Croatia}
\address[15]{University of Washington, Box 351560, Seattle, WA 98195, USA}
\cortext[cor2]{Corresponding author: carlo.gabbanini@ino.it}

%\mail{e-mail: xx}

\begin{abstract}
\noindent \normalsize
The new muon $(g$\,$-$\,$2)$ experiment E989 at Fermilab will be equipped with a laser calibration system for all the 1296 channels of the calorimeters. An integrating sphere and an alternative system based on an engineered diffuser have been considered as possible light distributors for the experiment. We present here a detailed comparison of the two based on temporal response, spatial uniformity, transmittance and time stability.  

\end{abstract}


\end{frontmatter}

% main text

%\maketitle
%\sloppy
%\twocolumn
%\setlength{\columnsep}{-5 mm}
%\oddsidemargin -.5cm
%\evensidemargin -.5 cm
%\textwidth 15cm
%\headheight 1.0in
%\topmargin -4cm
%\textheight 22cm

\section{Introduction}
\label{intro}
A new $(g$\,$-$\,$2)$ experiment, E989~\cite{carey09} at Fermilab, has been proposed to measure the muon anomalous magnetic moment to a precision of $1.6 \times 10^{-10}$ (0.14 ppm). To achieve a statistical uncertainty of 0.1 ppm, the total data set must contain more than
$1.5\times 10^{11}$ detected positrons with energy greater than 1.8 GeV.
In particular, the experiment will require a continuous monitoring and re-calibration of the detectors, whose response may vary on both a short timescale of a single beam fill, and a long one of accumulated data over a period of more than one year.

It is estimated that the detector response must be calibrated with relative accuracy at sub-per mil level to achieve the goal of the E989 experiment to keep systematics contributions due to gain fluctuations at the sub-per mil level on the beam fill scale (0-700 $\mu$s).  Over the longer data collection period the goal is to keep systematics contributions due to gain fluctuations at the sub-percent level. This is a challenge for the design of the calibration system because the desired accuracy is at least one order of magnitude higher than that of all other existing, or adopted in the past, calibration systems for calorimetry in particle physics~\cite{anfreville07, viret10}.

As almost 1300 channels must be kept calibrated during data taking, the proposed solution is based on the method of sending simultaneous light calibration pulses onto the silicon photomultiplier (SiPM) photo-detectors, through the active crystals ( made of PbF$_{2}$)  that make up the calorimeter. Light pulses should be stable in intensity and timing in order to correct systematic effects due to drifts in the response of the crystal readout devices. A suitable photo-detector system must also be included in the calibration architecture to monitor any fluctuation in time of the light source intensity and beam pointing as well as any fluctuation of the transmitted light along the optical path of the light distribution system, which could occur due to mechanical vibrations or to aging of the optics.

Some guidelines are defined to select the light source(s) and to design the geometry of the light distribution and monitoring; the following criteria are adopted to select the light source type:
\begin{itemize}
\item light wavelength must be in the spectral range accepted by the detector and determined by the convolution of the spectral density of the Cherenkov signal produced by electrons in PbF$_{2}$ crystals with the spectral transmission of the crystals, and with the photo-detection-efficiency (PDE) of the photo-detector; PDE is peaked around 420 nm for SiPMs.
\item the luminous energy of the calibration pulses must be in the same range of that produced by the conversion of the electron energy in the crystals, typically 1-2 GeV, into light ~\cite{fienberg15}. This corresponds to a luminous energy on each tower of a calorimeter station of  about 0.01~pJ, or to about 0.013~nJ for simultaneous excitation of all calorimeter readout channels (1300). The numbers quoted above are merely indicative of the order of magnitude and are derived by assuming that the readout of each crystal will produce about 1.5 photo-electrons per MeV with 35\% PDE for SiPMs and with 23\% coverage of the crystal readout face.
\item the pulse shape and time width must be suitable to infer on the readout capability in pile-up event discrimination: pulse rise/trailing time should be of the order of some hundred of picoseconds, the total pulse width should be of the order of 1~ns. However, as the SiPM detectors in the planned experiment give temporal tails of several ns, the time width requirements are less stringent.
\item the pulse repetition rate must be of the order of 10~kHz; this value will be tuned to obtain the best compromise between the need of having enough calibration statistics in the time interval when the maximum rate is achieved in the readout devices and the need to avoid pileup.

\end{itemize}

Among many different types of pulsed lasers commercially available, pulsed diode lasers in the blue seem to best address all the criteria listed above and are considered as a source for the calibration pulses. The light pulses will be distributed to the calorimeters in a way that will be defined after the completion of all tests required to qualify, in terms of light transmission and time stability,  all other optical elements of the calibration system. 

A particulary appealing candidate for the distributor is the integrating sphere, which offers  a high degree of output uniformity at the price of a rather large intensity loss. In the planned design, the laser pulses should be sent into one or more integrating spheres and then coupled into bundles of optical fibers. Each fiber will finally convey the light pulses into the crystals and SiPMs to be calibrated. As an alternative, we tested a system based on an engineered diffuser, having higher transmission efficiency but a lower degree of output uniformity.
In this paper we describe both tests.

\section{Integrating sphere and diffuser}

Integrating spheres with diffusely reflective white walls, also known as Ulbricht spheres, have been used for more than a century for the characterization of light sources, detectors, and for other photometric studies. We will not detail here their working principles. For some recent applications see for instance refs.~\cite{poikonen10, park13}, while for the theory see ref.~\cite{goebel67, sphere00}. Briefly, an integrating sphere is used to spatially integrate a radiant flux that can either be introduced into it by some input port or directly produced inside the sphere by a source. In either case, the light reflects diffusively several times inside the sphere so that, on some output port, the radiant flux can be considered uniform and isotropic.\\

We compared the performances of the integrating sphere with a distributor based on a engineered diffuser described in next section. \\

For the present scope, several measurements have been performed to test various characteristics of the integrating sphere and of the diffuser, i.e. the temporal response to short laser pulses, the spatial uniformity, the transmittance into a fiber bundle and the transmission stability over relatively long lapse of time.\\ 

\subsection{Temporal response  
\label{ssec:timeresp}}

Most integrating spheres are used as steady state devices. The standard analysis of their performance and application assumes that the light levels within the sphere have been constant for a time lapse long enough so that all transient response has disappeared. If rapidly varying light signals, such as short laser pulses are introduced into an integrating sphere, the output signals may be noticeably distorted by the “pulse stretching” caused by multiple diffuse reflections. The shape of the output signal is determined by convolving the input signal with the time response of the integrating sphere. The time response is of the exponential form $e^{-t/\tau}$, where the time constant $\tau$ is given by:

\begin{equation}
  \tau = -\frac{2}{3}	 \frac{D}{c} \frac{1}{\ln{{\rho}_{\rm eff}}}
\label{eq:tau}	
\end{equation}
where $\rho_{\rm eff}$ is the average wall reflectance, $c$ is the velocity of light and $D$ is the diameter of the integrating sphere~\cite{sphere00}. The average wall reflectance takes also into account the presence of input and output ports according to 

\begin{equation}
  \rho_{\rm eff} = \rho (S_{\rm tot}-S_{\rm open} ) /  S_{\rm tot}
\label{eq:rho_eff}	
\end{equation}where $\rho$ is the wall reflectivity, $S_{\rm tot}$ and $S_{\rm open}$ are the total and open surfaces of the sphere respectively.

\begin{figure}[h] \center
\includegraphics[width= 0.45\textwidth]{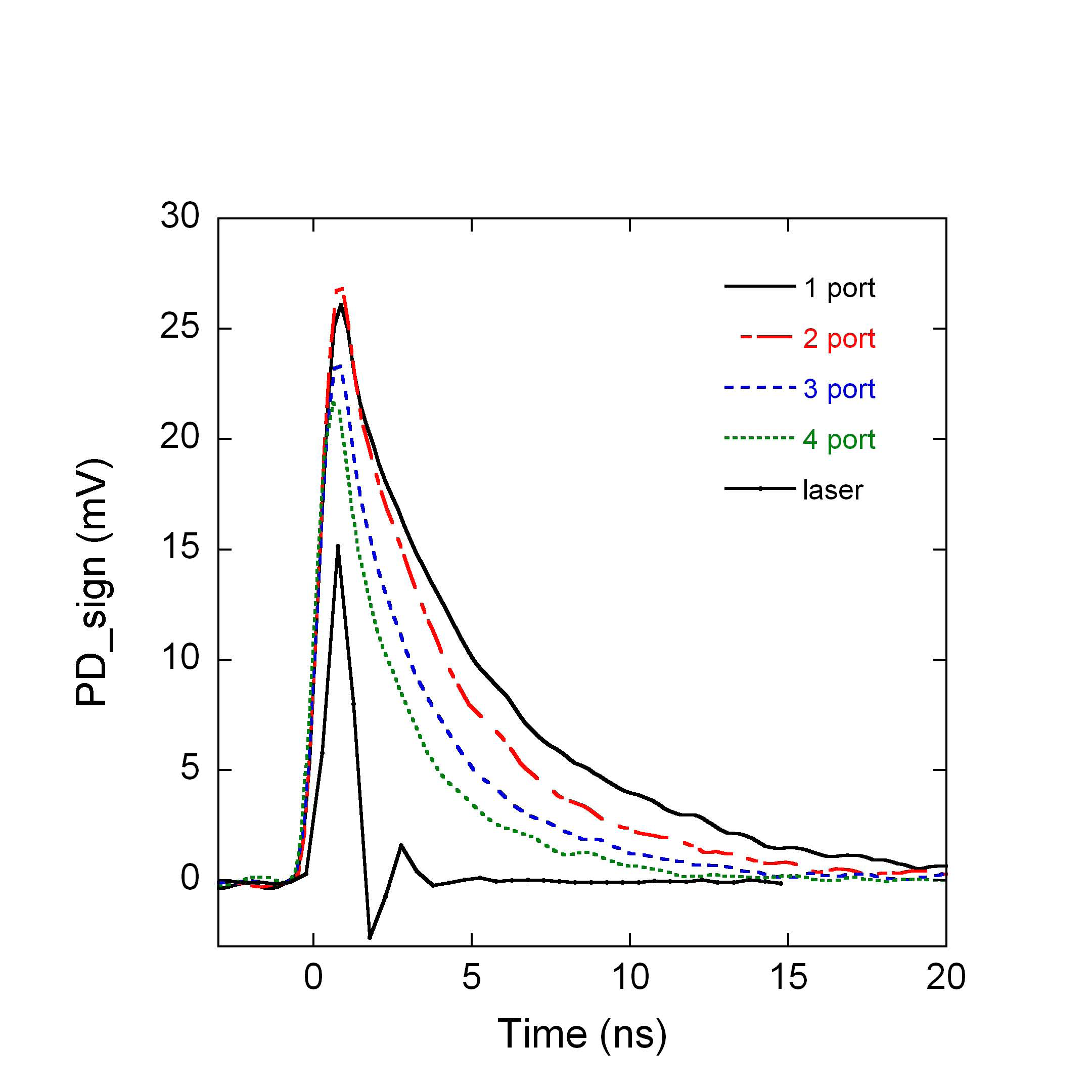}
\caption{(color online) Temporal response of the sphere to a short laser pulse under different conditions (see legenda).  The laser pulse measured directly on the PD has oscillations due to imperfect impedence matching.}
\label{fg:temporal}
\end{figure}

In our experiment we used an integrating sphere with a diameter of 2 inches (Thorlabs IS200-4). It has an inner surface reflectivity $\rho$ of 99~\% in the visible range, four 1/2~inch input/output ports and a 3~mm diameter, recessed detector opening. To study the temporal response of the sphere to short light pulses,  a frequency doubled Ti-sapphire laser (wavelength 400~nm) with pulse duration of the order of 300~fs and and initial pulse energy of about 1~mJ, attenuated to a final value of some $\mu$J, has been used. This is not the laser that will be used for the E989 experiment, but it is useful in this case for its very short pulse duration, that can be considered as a delta function on our timescale. For detection we used a Hamamatsu model S5973-02 PIN photodiode (PD) with 1.5~GHz bandwidth and high sensitivity in the blue region (0.3~A/W, QE = 91\% at $\lambda$~=~410~nm).
The photodiode was connected directly to a fast oscilloscope (LeCroy Wavesurfer 452, 500~MHz bandwidth, 2~Gs/s sampling rate). 

\begin{table}[width= 0.5]
	%\center{ 
	\begin{tabular}{|l|l|l|} \hline
	open ports & calc.value (ns) & meas.value (ns) \\ \hline
	1 & 4.2 $^{+2.1}_{-0.5}$ & 4.6 $\pm 0.1$ \\ \hline
	2 & 2.6 $^{+0.7}_{-0.4}$ & 3.3 $\pm 0.1$ \\ \hline
	3 & 1.9 $^{+0.4}_{-0.1}$ & 2.4 $\pm 0.1$ \\ \hline
	4 & 1.5 $^{+0.2}_{-0.1}$ & 1.7 $\pm 0.1$ \\ \hline
	\end{tabular}
	\caption{Calculated vs experimentally measured decay constant of PD signals on the sphere (after deconvolution) as a function of the number of open ports.}
	\label{tab:decay}
\end{table}
 
First the signal was recorded from the PD directly positioned on the path of the laser beam after attenuation. By considering a Gaussian pulse, the full width at half maximum (FWHM) is 0.8~ns, giving the effective bandwidth of the electronics. Then, the temporal response of the sphere was measured in 4 different conditions, i.e. with the PD in the small detector opening and with 1, 2, 3 or 4 ports of the sphere successively open (one port is always used as laser beam input). The experimental results are shown in Fig~\ref{fg:temporal}. 
 While the rise time results to be about 1~ns, the decay time has different time constants, going from about 5~ns in the first case to about 2~ns with all ports open. The time constants are derived by deconvolving the signals with the response function of the electronics. The results are summarized in Table~\ref{tab:decay}, together with what expected from the time constant formula~\ref{eq:tau}.  The measurements have different effective reflectivity because the total reflective surface decreases when ports are opened.  The error in the calculated decay constant is due to an indetermination of 0.4~\% on the sphere reflectivity and a maximum estimated 30~\% reflectivity of the detector port, due to the PIN diode, its can and its aluminum holder. \\
We compared the sphere response with that of an engineered diffuser,  produced by RPC Photonics and marketed by Thorlabs (mod. ED1-C20), consisting of structured microlens arrays that transform a Gaussian input  beam into a flat top one~\cite{sales03}. The input beam should be collimated and expanded to cover a high number of microlenses while the output beam has a divergence of 20 degrees. All optical elements, i.e. the telescope, the diffuser and the fiber bundle (see further sections), are mounted on a single cage. For the diffuser we observed that there isn't any variation of the temporal profile of the laser pulse.  

\subsection{Spatial uniformity  
\label{ssec:unif}}
It is well known that the integrating sphere ensures a high uniformity of the radiation emitted at an output window~\cite{park13}. However this is generally true when all transient response has disappeared. We investigated the luminance uniformity in the case of a single laser pulse. In order to measure the uniformity at an output port, we used a Single-Lens Reflex (SLR) camera, that allows manual adjustments of the shooting parameters. Because of the optical system and the distance of the sensor from the window, the camera collects radiation emitted by a specific portion of the sphere. This is somewhat analogous to what occurs in collecting radiation from the port of the sphere by means of an optical fiber (or a bundle of fibers), which would be the case assumed in the $(g$\,$-$\,$2)$ experiment. 

\begin{figure} [h] \center
\includegraphics[width= 0.48\textwidth]{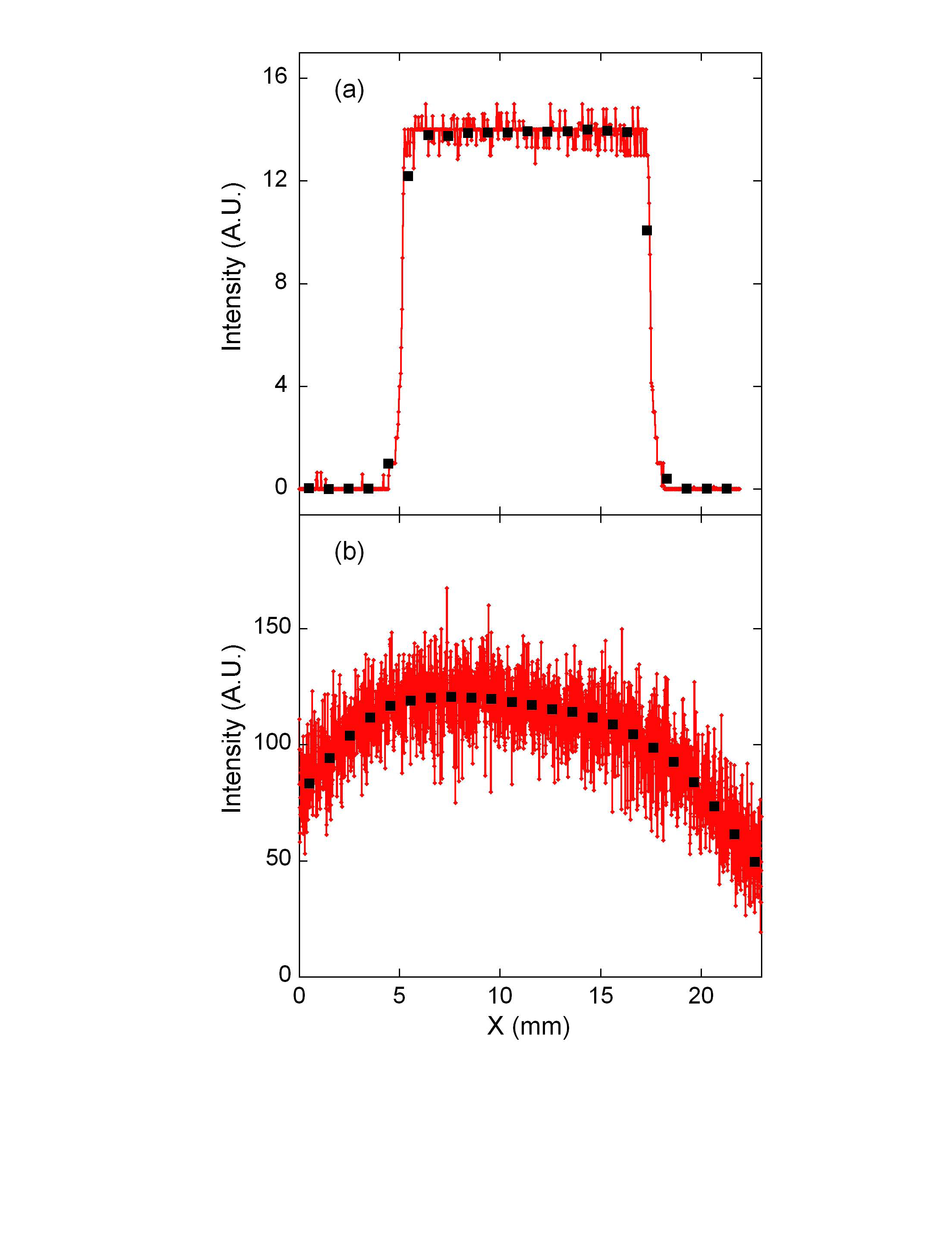}
\caption{Experimental luminance profiles of a single pixel row (daggers) and averaged 900 pixels (squares). Graph a) is for the sphere and b) for the diffuser. \label{fg:unif}}
\end{figure}

Images have been recorded with a  Pentax camera (model K100D), with an exposure time small enough (1/30~s) to acquire a single laser pulse of a frequency doubled Nd:YAG with pulse width 10~ns and 20~Hz repetition rate (Continuum I-20). As we kept the signal weak enough to avoid saturation, the resulting intensity fluctuations were quite significant. In order to reduce the error and to estimate the signal collected by an optical fiber with 1~mm diameter, we averaged along a strip, 30 pixels wide, corresponding to 1~mm. Furthermore, the averaged line has been divided into sections of 30 pixels, further averaged, so that each point is the average of 900 pixels contained in a square of 30 pixels side. In Fig.~\ref{fg:unif}a  the comparison between the single row data and the data averaged over 900 pixels are shown. The uniformity reaches a value of 0.99 in the 8 mm central part.\\
	Images have been recorded also for the diffuser. In this case the laser beam is expanded by a telescope to a diameter above 1 cm before impinging into the diffuser. The camera is placed just after the diffuser without the lens, so that its CCD sensor is about 5 cm away from the diffuser surface. The analysis of the image of a single laser pulse image is shown in  Fig.~\ref{fg:unif}b. The uniformity is in this case 0.97  in the 8 mm central part.

\subsection{Transmission measurement \label{ssec:trans}} 
As the system should operate as a light distributor through optical fibers in the $(g$\,$-$\,$2)$ experiment, we measured the transmission of two combined systems: sphere+fiber bundle and diffuser+fiber bundle. In order to make reliable measurements we used a fraction (150~mW) of a high-power laser with continuous emission (frequency-doubled Nd:YAG laser, wavelength 532~nm, P$_{max}$ =7~W, model Coherent VERDI), which has an excellent output power stability. We used a power meter with two interchangeable heads with different full scale to detect the signals.
In the first measurement, a fiber quartz  bundle (60 fibers of diameter 200~$\mu$m, manufactured by Leoni Fiber Optics Inc.) was  connected to the sphere using an adapter. The output power from each fiber was measured using the power meter with the lower full scale head. This measurement has been carried out for two ports of the sphere at 90 degrees from each other that are called the 90 degree port and the north pole port. 
 
\begin{figure}[h]  \center
%\vspace{2in}
\includegraphics[width= 0.48\textwidth]{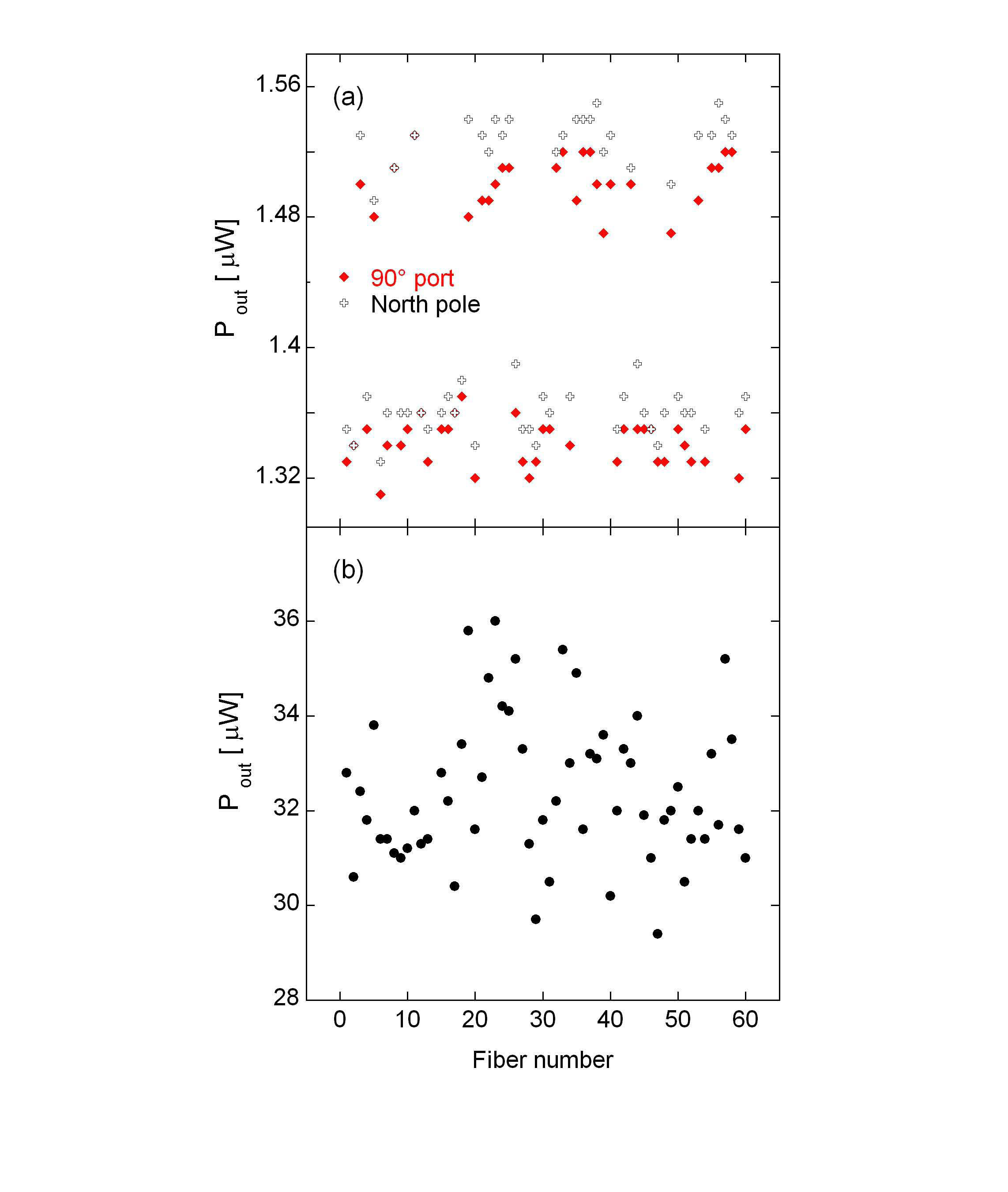}
\caption{(color online) Transmitted power from each fiber of the bundle. Graph a) is for the sphere;  diamonds correspond to the bundle connected to the 90° port while crosses to the bundle connected to the north pole. Graph b) is for the diffuser with focusing lens.}
\label{fg:trans}
\end{figure}

 Fig.~\ref{fg:trans}a clearly shows that the fibers are grouped into two classes with different transmission. Such behavior is attributed to the bundle, not to the sphere, as by changing the port of the sphere to which the bundle is connected, the qualitative behavior does not change. From a quantitative point of view, the north pole port has a slightly higher output transmission (about 2~\%). This can be attributed to the fact that the surface of the sphere that faces the 90 degrees port has the 3~mm diameter hole of the detector port, while the surface facing to the north pole port is all reflective.

\begin{figure}[h] \center
%\vspace{2in}
\includegraphics[width= 0.45\textwidth]{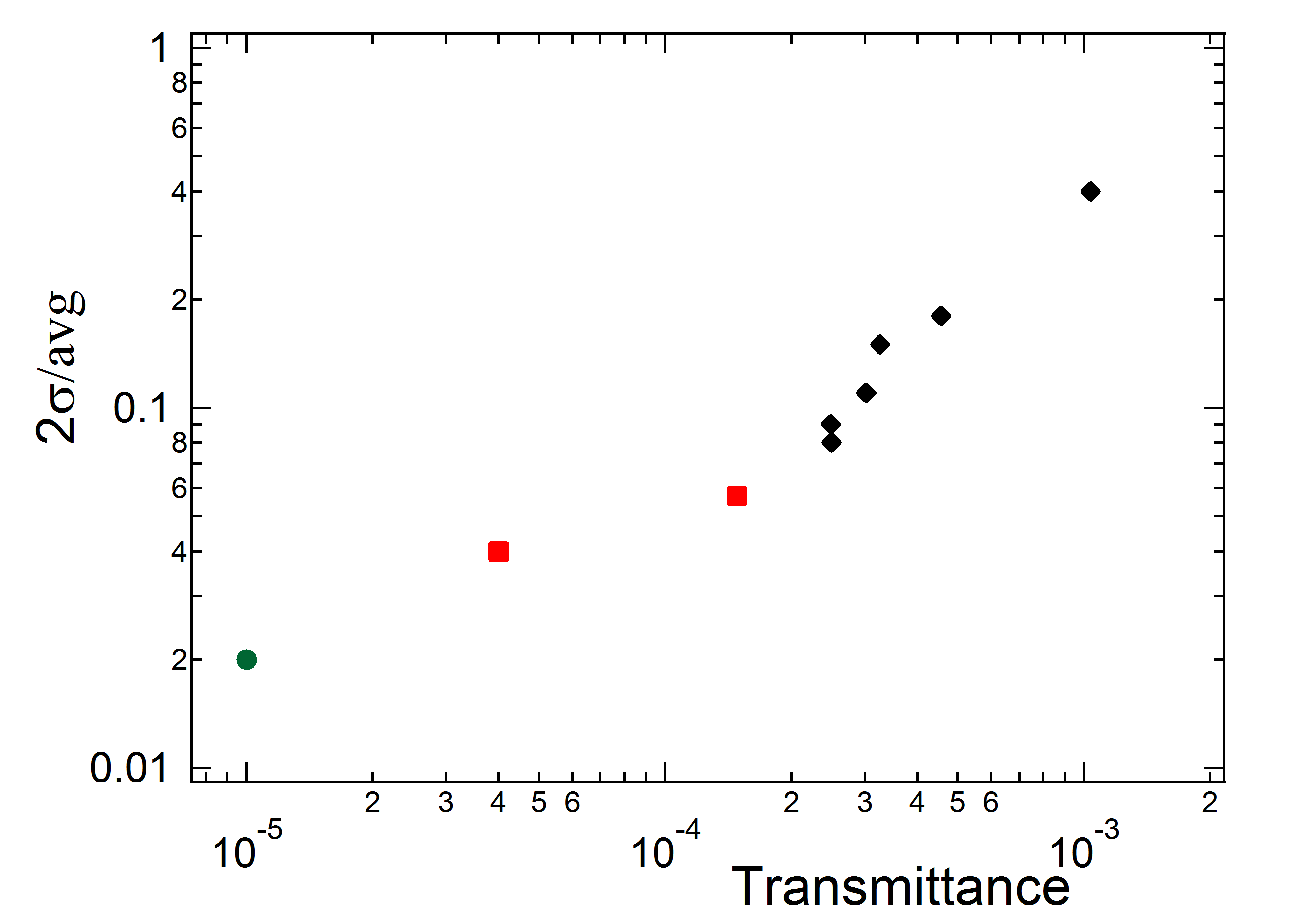}
\caption{(color online) Inhomogeneities of the transmitted powers of the fibers as a function of the averaged transmission coefficient. Circle: sphere; squares: diffuser without focusing lens; diamonds: diffuser with focusing lens.}
\label{fg:disom}
\end{figure}

%Fig.~\ref{fg:trans1} shows the histogram of the occurrences at various power levels. Two distributions of Gaussian type are visible, for  both  sphere ports. 

The measured overall transmission coefficient (sphere + bundle) is thus $T = (1.0 \pm 0.1)\times 10^{-5}$. The same transmission, with a larger indetermination, has been obtained also using a pulsed laser and a fast PIN diode.\\
We can compare this experimental value with that expected from theory. The sphere surface radiance, $L_S$, of the sphere for an input flux $\Phi_i$, is given by:

\begin{equation}
 L_S=\frac{\Phi_i}{\pi A_{sphere}} \frac{\rho}{1-\rho (1-f)}
\label{eq:radiance}	
\end{equation}
 
where $\rho$ is the surface reflectance, $A_{sphere}$ the sphere surface and $f$ is the port fraction ($f$ = $A_{open}$ /$A_{sphere}$). 
Due to the numerical aperture (NA) and losses at the air/fiber interface,  the light coupling efficiency of an optical fiber is greatly reduced: 

\begin{equation}
 \Phi_{out}= \pi L_S A_f (1-R) (NA)^2 \approx 0.037 \pi L_S A_f
\label{eq:coupling}	
\end{equation}

where $A_f$ is the surface of the fiber, $R$ is the reflectance at the fiber face (e.g. 8~\%) and NA is the numerical aperture. 
The overall transmission coefficient for the complete system (sphere + fiber bundle) results:

\begin{equation}
 \frac{\Phi_{out}}{\Phi_{i}} = 0.037 \frac{A_f}{A_{sphere}} \frac{\rho}{1-\rho (1-f)} = 1.8 \, 10^{-5}
\label{eq:trans}	
\end{equation}

assuming the fiber diameter equal to 200 $\mu$m, NA= 0.2, $\rho$ = 0.99 and  $f$ = 0.0215. 
This calculated value is an upper limit for the fiber output, because it is necessary to take into account the length, the material extinction coefficient and the exit interface reflection of the fiber.\\

These measurements have been done also for the diffuser. As the output beam from the diffuser diverges, by changing the distance between the bundle and the diffuser we could collect more or less light. For some measurements we added a lens after the diffuser to partially focus the light into the fiber bundle and collect more light.  Fig.~\ref{fg:trans}b  shows the transmitted powers of the fibers in one position of the focusing lens.  Here the power values are more inhomogeneous than for the sphere, as expected, and the two classes of fibers with different transmission are not distinguishable. By varying the geometry in order to collect more or less light,  we observed that the inhomogeneities, that we consider as twice the standard deviation divided by the average transmitted power,  increase with the transmission coefficient, as shown in  Fig.~\ref{fg:disom}. The values are computed considering just one of the two classes of fibers with different transmission. The transmission can exceed  $T=10^{-3}$ for each fiber, but at the expense of a large inhomogeneity. In this figure also the value for the sphere is reported. While for the sphere there is no possibility to adjust the transmission coefficient, for the diffuser one can establish a uniformity goal, for instance 10~\%, and adjust it suitably.
\\
The transmission can be significantly increased both for the sphere and the diffuser by using fibers with higher diameter or higher numerical aperture. For our distribution purposes we plan to use 1mm diameter PMMA fiber, which should allow nearly a factor 25 gain in the transmission factor.

\subsection{Intensity  stability over time
\label{ssec:timestab}}

\begin{figure}[h] \center
\includegraphics[width= 0.45\textwidth]{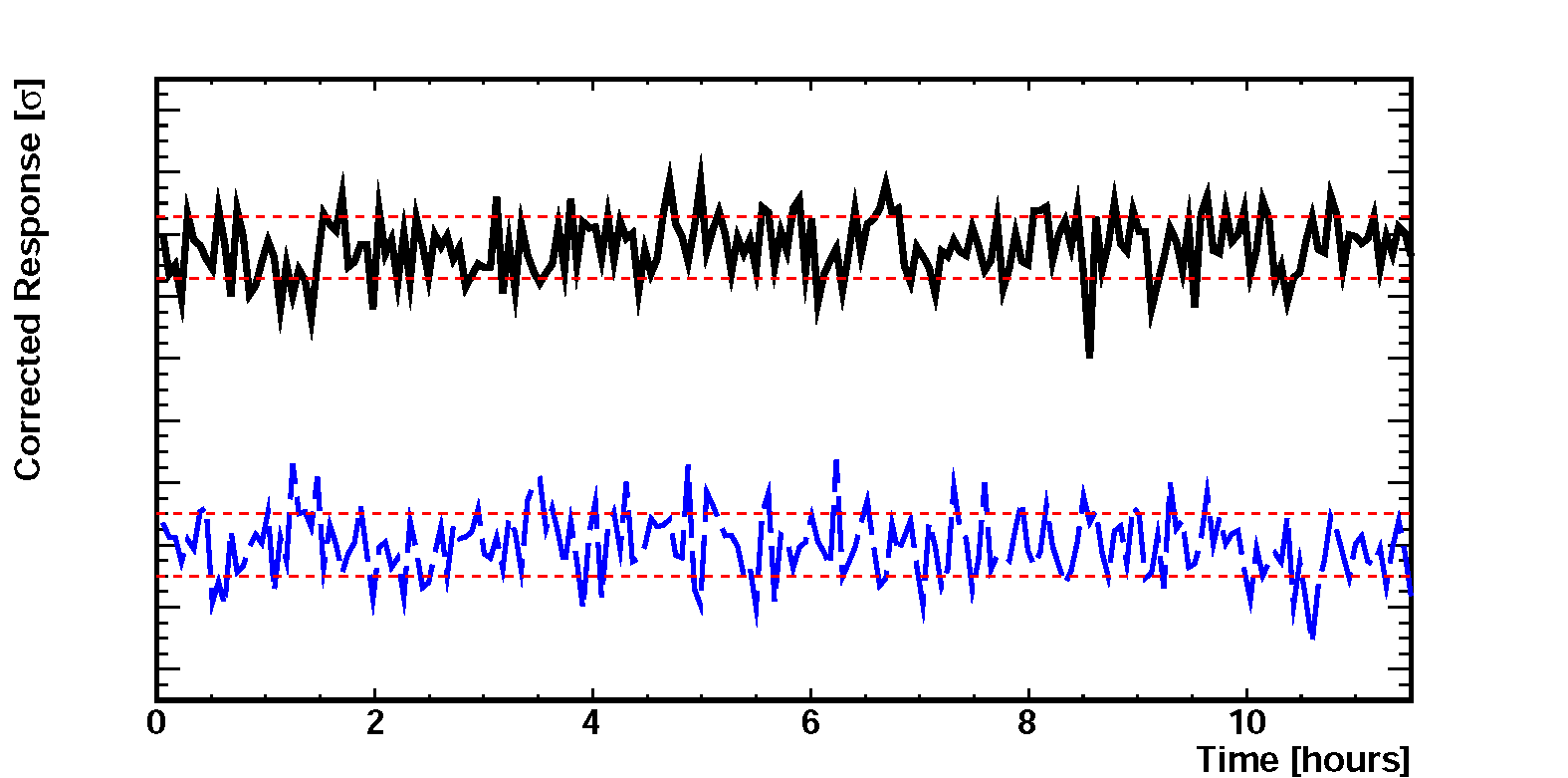}
\caption{Sphere case. Upper line: SiPM response's deviation from its initial value as a function of time in units of the standard error on the mean; lower dashed 
trace: simulated random Gaussian fluctuations; horizontal dashed lines: $\pm1$  $\sigma$ deviations from the mean value over the run.\label{fg:stab1}}
\end{figure}

\begin{figure}[h] \center
\includegraphics[width= 0.45\textwidth]{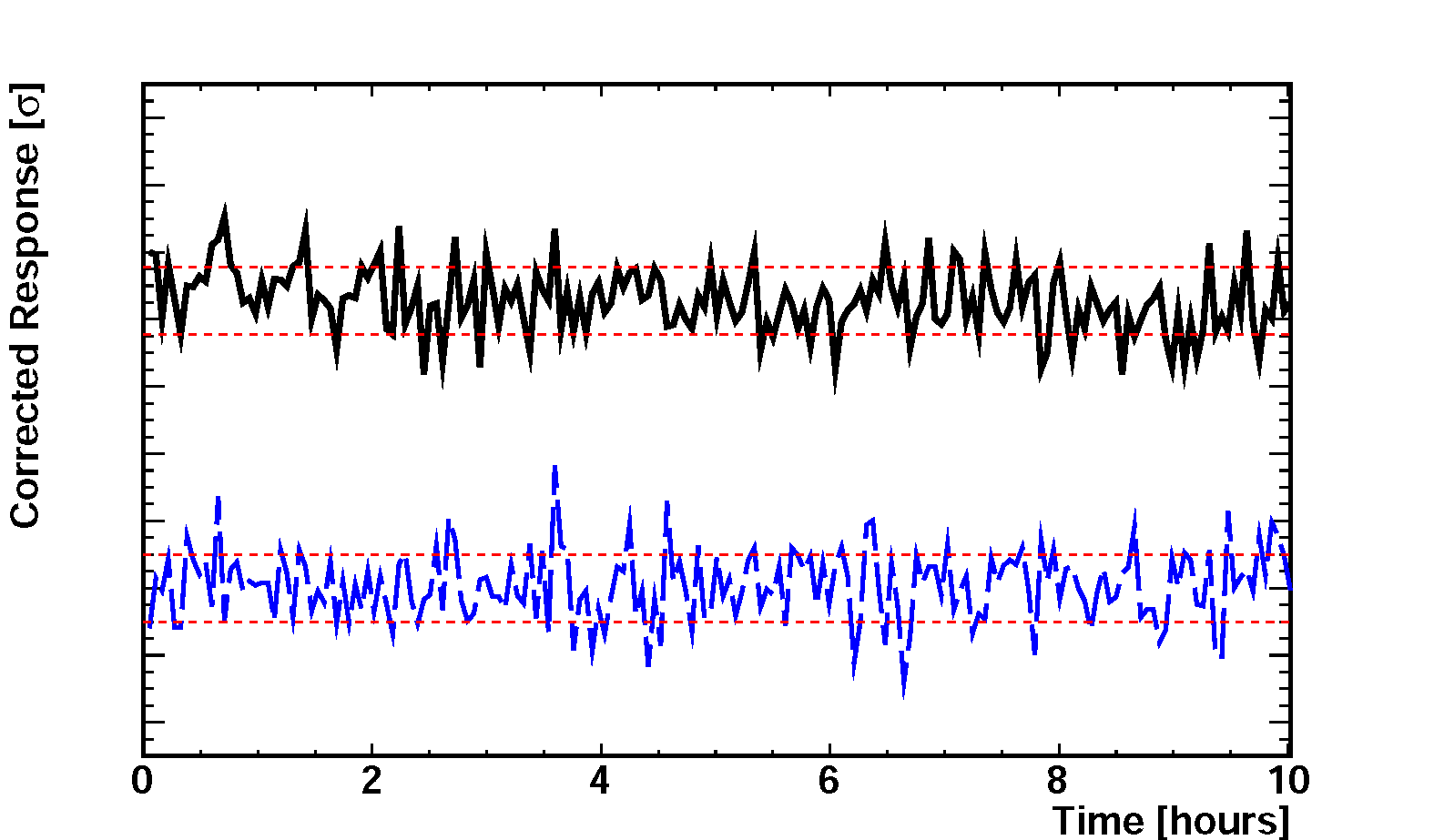}
\caption{Diffuser case. Upper line: SiPM response's deviation from its initial value as a function of time in units of the standard error on the mean; lower dashed trace: simulated random Gaussian fluctuations; horizontal dashed lines: $\pm1$  $\sigma$ deviations from the mean value over the run.\label{fg:stab2}}
\end{figure}

Another very important test parameter for the distribution system of $(g$\,$-$\,$2)$ is the long-term stability of the signals conveyed to the silicon photomultipliers in the calorimeters. Fluctuations between the signals sent to different channels, not  monitored at the end of the distribution system, result in a calibration systematic error of SipM gains. 
To assess the stability we carried out some experiments, measuring the time variation of the light output of the fibers of a bundle connected either to the integrating sphere or to the diffuser. The measurements have been done using a setup similar to the experiment reported in ~\cite{fienberg15}. A PicoQuant LDH-P-C-405M pulsed diode laser, having a pulse width of about 600 ps at a wavelength of 405 nm, illuminated either the sphere or the diffuser, with a fiber bundle as  light extractor. The fibers of the bundle, produced by Vinci, are in optical silica and have a 0.6 mm diameter and a numerical aperture of 0.39. Twelve of the fibers were fixed by SMA connectors to a panel facing an array of twelve PbF$_2$ crystals. The light was collimated by aspheric lenses before entering the crystals, and detected at the end face of the crystals by large area SiPMs. The light arriving to the SiPMs corresponded to about 750 photoelectrons for the sphere and a factor two more for the diffuser. A neutral filter was used in the last case to attenuate the light in order to avoid saturation of the SiPMs. The laser pulses, having a repetition rate of a few kHz, were detected and digitized along several hours and subsequently analysed. 
The fitting procedure of the pulses is described in ~\cite{fienberg15}.
\\
In figure~\ref{fg:stab1} the upper line shows a SiPM response's deviation from its initial value as a function of time in units of the standard error on the mean,
measured from the first 1000 samples, for the sphere case. The points included in the plots are averages of 1000 laser shots. SiPM responses are corrected by the mean of all twelve SiPMs to eliminate the effects of temperature, bias, and laser drifts. The dashed lines denote $\pm1$  $\sigma$ deviations from the mean value over the run. The lower trace shows
what comes from completely random Gaussian fluctuations, such as expected from photostatistics. The same analysis done for the diffuser measurements is shown in  figure~\ref{fg:stab2}. In this case the standard error on the mean is different from that of the sphere because of the increased intensity, and consistent with the change due to photostatistics.
\\
We can observe that for both measurements the fluctuations are limited by SiPM photostatistics at the level of 0.1~\%, meaning that the systematic fluctuations of both distribution systems are below this level.

\section{Conclusions}

In this paper we studied some characteristics of two devices, i.e. an integrating sphere and an engineered diffuser, to be used, together with an optical fiber bundle, as  possible light distribution systems. They should be part of the calibration system of the forthcoming E989 $(g$\,$-$\,$2)$ experiment at Fermilab, where  short ($\approx$~ns) laser pulses will be delivered  to  a large number (1300) photodetectors for an extended period of time.  Therefore we tested in particular the sphere's and the diffuser's temporal response, their illumination uniformity when subject to short laser pulses, their trasmission capabilities when coupled with an optical fiber bundle, and finally their intensity stability over a relatively long period of time.\\
For the sphere our results show that a short laser pulse produces spatial illumination fluctuations in the output ports below 0.2~\%, when integrated over windows of 1 squared mm area (i.e. of the order of the optical fiber surface). This is obtained at the price of a temporal lengthening of the laser pulse up to a few ns, depending on the number of open ports of the sphere. This may not be a problem for the $(g$\,$-$\,$2)$ experiment, as the pulse width remains below the SiPM response time~\cite{fienberg15}.  Laser pulses are also severely attenuated by the sphere when used to couple light into a fiber bundle, resulting in a measured trasmission of the order of $10^{-5}$ in a single fiber of 200~$\mu$m diameter and an estimated one of $2.5 \times 10^{-4}$ into a 1~mm one.
Nevertheless, such a system  is able to uniformly split a single laser pulse into more than 100 optical fibers simultaneously, with a good long-term mechanical stability, and an estimated total transmission coefficient of the order of  2~\%. \\
The test has  also shown that it is possible to achieve even higher transmission coefficient for each fiber when using a system based on an engineered diffuser, at the expense of a lower uniformity. We expect also a slightly worse long-term stability for the diffuser, but this is not evident at the measured level.
The diffuser has the advantage of leaving unaltered the temporal profile of the laser. \\
The final choice  for the light distribution system of the $(g$\,$-$\,$2)$ experiment calibration will be done by considering several issues, including the total required laser power.\\

%BeginExpansion

{\bf Acknowledgements}
\\
This research was supported by Istituto Nazionale di Fisica Nucleare (Italy). We thank F.Scuri and S.Veronesi for useful discussions and A.Barbini and F.Pardini for technical assistance.
%EndExpansion
\\
\\

%\textbf{References}
%\begin{thebibliography}{00}

\footnotesize{
\bibliography{sfera} %your .bib file
\bibliographystyle{plainnat} %the RSC's .bst file
}

%\end{thebibliography}

\end{document}